\documentstyle[twocolumn,aps,prl]{revtex}
\begin{document}
\draft
\title{Interference of resonances and the double-pole of a $S$-matrix in a double well system}
\author{Wim Vanroose}
\address{Departement Wiskunde-Informatica, Universiteit Antwerpen\\
Groenenborgerlaan 171, 2020 Antwerpen, Belgium\\
vanroose@ruca.ua.ac.be}
\date{\today}
\maketitle

\begin{abstract}
We discuss the interaction between two resonant states in a quantum double-well structure. The behaviour of the resonant states depends on the coupling between the wells, i.e. the height and width of the barrier that separates them. We distinguish a region with resonant tunneling and a region where the
two resonances repel each other. The transition between the two regions is marked by a double pole of the $S$-matrix.
\end{abstract}
\narrowtext
{\it Introduction} --- Lately the interference effects of resonances and in particular the occurrence of a double pole of the $S$-matrix has been examined frequently \cite{latinne,kylstra,vanroose,joachain,mag}. Several, widely different systems where double poles can occur have been identified, such as: atomic states in intense laser fields, general two channel systems, and other systems.  Recently Hernandez {\it et al}\cite{mond} have investigated a model with two spherical cavities bounded by $\delta$-function barriers and shown that a double pole of the $S$-matrix can be induced by tuning the parameters of the model. In this letter, we study the interference effects between two resonances in a one-dimensional quantum system with a double square well. In our opinion it is, besides being more general than a model with singular potentials, more suitable for a physical discussion about the genesis of the double pole and its relation to the internal structure of the system, as it can be related to concepts like coupling and tunneling.  Such a double well structure is similar to the structure of two vertically coupled quantum dots or an ``artificial molecule'' \cite{tarucha}. To our knowledge (and surprise), these aspects of the double well system have not been described in the literature.

As model problem we take the motion of a particle in a one-dimensional potential $V(x)$ governed by the Schr\"{o}dinger equation 
\begin{equation}
-\frac{1}{2}\frac{d^{2}}{dx^{2}}\psi (x)+V(x)\psi (x)=E\psi ,
\label{schrodinger equation}
\end{equation}
in units where $m=1$ and $\hbar=1$. The potential $V(x)$ has the form of a double well as shown in figure 1. Between the infinitely repulsive wall at $x=0$ and the free space at $x>x_{4}$ where $V=0,$ there are two square wells separated by two square barriers.

The aim of our calculation is the study of resonances of this system by inspection of the poles of the $S$-matrix and to show how a double pole is formed and to discuss how it marks the transition between two different scattering regimes.

The model contains eight parameters, i.e. the positions $x_{i}(i=1,2,3,4)$ and heights $V_{i}(i=1,2,3,4)$ of the four discontinuities of the potential. We have chosen here to fix six of the parameters ($V_{1},$ $V_{2},$ $V_{4},$ $x_{1},$ $x_{3}-x_{2},$ $x_{4}-x_{3}$) and to vary the depth of the outer well $V_{3}=V$ and the thickness of the inner barrier $x_{2}-x_{1}=D$ .

We start by looking at the interaction between a resonant state that resides in the first well and a resonant state that resides in the second well. We are especially interested in how this interaction depends on the parameter $D$,  the distance between the wells (quantum dots), which measures the coupling strength between the two wells. Therefore, we study how the lifetime and position of the resonant states evolves when $D$ changes.

The Schr\"{o}dinger equation (\ref{schrodinger equation}) is solved exactly.
The solution in each segment $]x_{i},x_{i+1}]$ is 
\begin{equation}
\psi (x)=A_{i}\sin (K_{i}x+\delta _{i}(E))\quad \mbox{when}\quad x_{i}<x\leq %
x_{i+1},
\end{equation}
where the momentum $K_{i}$ is related to the potential $V_{i}$ in that particular segment of space 
\begin{equation}
K_{i}=\sqrt{2(E-V_{i})}
\end{equation}
The phase shifts $\delta _{i}(E)$ and $\delta _{i+1}(E)$ in two successive segments are related by the logarithmic derivative at the edge $x_{i+1}$ that separates the two segments. 
\begin{equation}
\frac{\tan (K_{i}x_{i+1}+\delta _{i}(E))}{K_{i}}=\frac{\tan
(K_{i+1}x_{i+1}+\delta _{i+1}(E))}{K_{i+1}},
\end{equation}
which results in the relation 
\begin{eqnarray}
&&\delta _{i+1}(E)=-K_{i+1}x_{i+1}  \nonumber \\
&+&\mbox{atan}(\frac{K_{i+1}}{K_{i}}\tan (K_{i}x_{i+1}+\delta _{i}(E)))
\label{matching condition}
\end{eqnarray}
Since the solution fits the zero boundary condition at the hard wall in $x=0$, we know that the phase $\delta _{0}$ of the solution in the first well is zero. By successive application of the matching condition (\ref{matching condition}), we find an exact expression for the phase shift $\delta (E)$. The $S$-matrix is 
\begin{equation}
S(E)=\frac{1+i\tan \delta (E)}{1-i\tan \delta (E)}.
\end{equation}
With this exact expression of the $S$-matrix we can locate the position of its complex poles 
\begin{equation}
E=E_{r}-i\Gamma /2.
\end{equation}
These poles are found by an algebraic computer package that searches for the zero minima of $\mbox{abs}(1-i\tan \delta (E))$ in the complex energy plane.

{\it Results} ---
The model parameters are now chosen in such a way that, for weak coupling between the wells, there is only one resonant state in each well and these states have nearly equal energy $E_{1}\approx E_{2}$ and vastly different width $\Gamma _{1} \ll \Gamma _{2}$. This can be done by choosing a large value of $D$, for which the wells are weakly coupled, and to adjust $V$ so that the system has one doublet of resonant states.\ So for $V_{1}=0,$ $V_{2}=4,$ $V_{4}=4,$ $x_{1}=1,$ $x_{3}-x_{2}=1,$ $x_{4}-x_{3}=0.3$, $D=2$ and $V=1.04$ we find two resonant states with complex energies $E_{1}=2.49-i0.00394$ and $E_{2}=2.43-i0.309$, i.e. the two resonances have nearly equal energy but totally different width. Resonance state $(E_{1}$, $\Gamma _{1})$ resides in the inner well and is narrow since it has to tunnel through two barriers to escape, whereas resonance state $(E_{2}$, $\Gamma _{2})$ resides in the outer well and is broad since it is confined by only a thin barrier. We then decrease the thickness $D$ of the barrier and study the behaviour of the two resonances. This is done by following the $S$-matrix poles in the complex
energy plane. The results are shown in figures \ref{trajectories} and figure \ref{width}. The figure \ref{trajectories} show the trajectories the $S$-matrix poles follow in the complex plane. Figure \ref{width} shows the individual dependence of $E_{r}$ and $\Gamma $ on $D$. We also consider the occupation probability of the two wells 
\[
P_{1}(E)=\int_{0}^{x_{1}}\mid \psi (E,x)\mid ^{2}dx
\]
and 
\[
P_{2}(E)=\int_{x_{2}}^{x_{3}}\mid \psi (E,x)\mid ^{2}dx
\]
They are shown in figure \ref{tunneling} for $D=0.5$ and $D=1.5$.

{\it Discussion} --- We distinguish two regions. For $D>1.1$, both complex poles attract each other, the resonance energies remaining almost equal. The narrow resonance broadens while the broad resonance gets narrower. This situation corresponds to resonant tunneling: the coupling of the two wells increases the tunneling probability out of the inner well. Meanwhile, it hinders the outward tunneling from the outer well, as the particle can now tunnel more easily into the inner well and make a detour before it goes to the continuum. From fig.5, we see that the occupation probabilities of the two wells are in anti-phase: the states are localized in the wells and at resonance either one of the wells is occupied.

For $D<1.1$, the coupling has become strong enough so that the two resonances are delocalized and belong to the double-well system as a whole. A thinner barrier will increase the coupling and force the two states to repel. One resonant state goes up in energy, the other down. The width of one increases because it has more energy to tunnel through the barriers; the width of the other gets smaller because it has less energy to break trough the barriers. Now the plot (fig.6) of the occupation probabilities shows a positive correlation: at resonance, both wells are occupied.

Finally, we arrive at a situation where $D=0$ and the barrier has vanished. We have now one broad well with two resonant states with totally different energies.

The trajectories change dramatically, when we slightly alter the bottom of the outer well to $V=1.03$. The trajectories of this system are displayed in fig.2b. Compared to fig.2a they interchange their identity beyond the transition point when $D>1.1$. From this we conclude that, at a critical value of the second parameter $V$, the trajectories at the transition point around $D=1.1$ will degenerate into a double pole.

The features of the mechanism of interfering resonant states in a double well system can also be seen if other parameters of the model are taken as variables, e.g. the height of the dividing barrier rather than its thickness and the depth of the inner well rather than of the outer well. If we should start with two resonant states that differ seriously in energy, the transition from ''resonant tunneling'' to ''level repulsion'' would not be so pronounced, i.e. the starting point of level repulsion would not be so clear. However, a transition will certainly take place. We speak of a transition region, where the coupling is intermediate.

{\it Conclusions} --- 
We have investigated the interference of resonances and the degeneracy into a  double pole of the $S$-matrix in the case of a double square well system. Because the exact analytical solution of the Schr\"{o}dinger equation is available, the scattering wave functions and the trajectories of $S$-matrix poles can be easily analyzed in detail.

We have shown that this simple, yet physically meaningful,  model contains all the relevant aspects of the interference mechanism.

We have demonstrated that the double pole marks the transition between two different scattering regimes: the resonant tunneling regime for weak coupling and the level repulsion regime for the strong coupling situation.

We suggest that this analysis is generic and can also be useful to classify scattering regimes in other systems with double poles. An example is a multi-channel system, where we identify a weak coupling regime where resonant states are described by a one-level approximation \cite{fano} and a strong-coupling regime where different resonant states interfere and a multi-level formula is necessary \cite{harris2,wang,wintgen} 

{\it Acknowledgements} --- 
The author is indebted to Prof. P. Van Leuven for many interesting discussions and acknowledges support from the ``Fonds voor Wetenschappelijk Onderzoek --- Vlaanderen''

\begin{figure}
\caption{The quantum double-well structure is such that we have two resonant
states, one in each well, with almost equal energy.}
\label{model}
\end{figure}

\begin{figure}
\caption{Trajectories of the $S$-matrix poles for decreasing distance $D$ between the wells. In the first figure is $V=1.04$, in the second $V=1.03$. The numbers in the figure indicate the thickness $D$. }
\label{trajectories}
\end{figure}

\begin{figure}
\caption{The position and width of the resonant states for each thickness of the barrier. The depth of the outer  well is $V=1.04$. One notices a sudden transition between the resonant tunneling and the level repulsion regime.}
\label{width}
\end{figure}

\begin{figure}
\caption{The energy dependency of the occupation probabilities $P_1(E)$ and $P_2(E)$. For a distance $D=1.5$ on top, for $D=0.5$ below.}
\label{tunneling}
\end{figure}

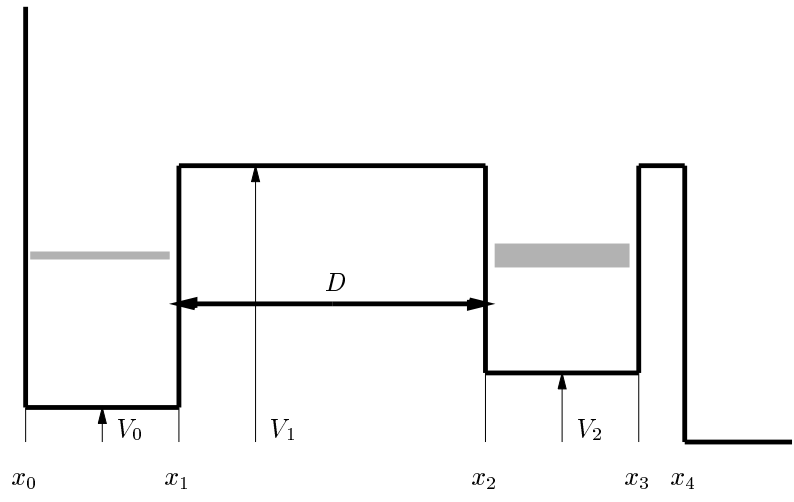
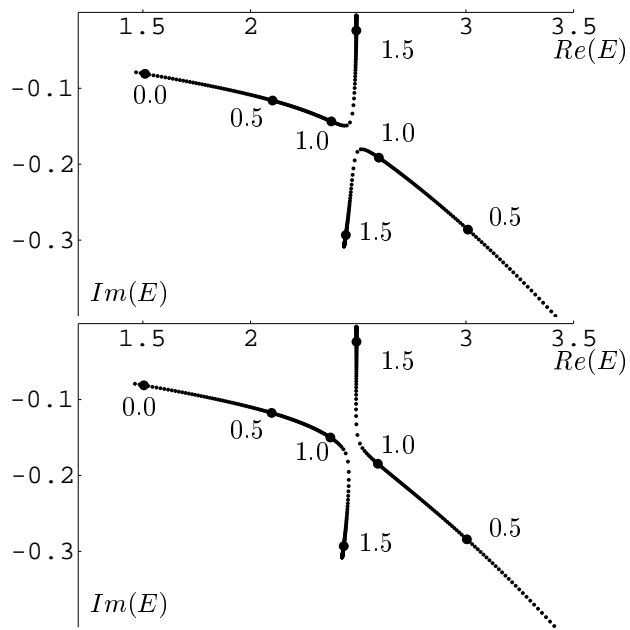
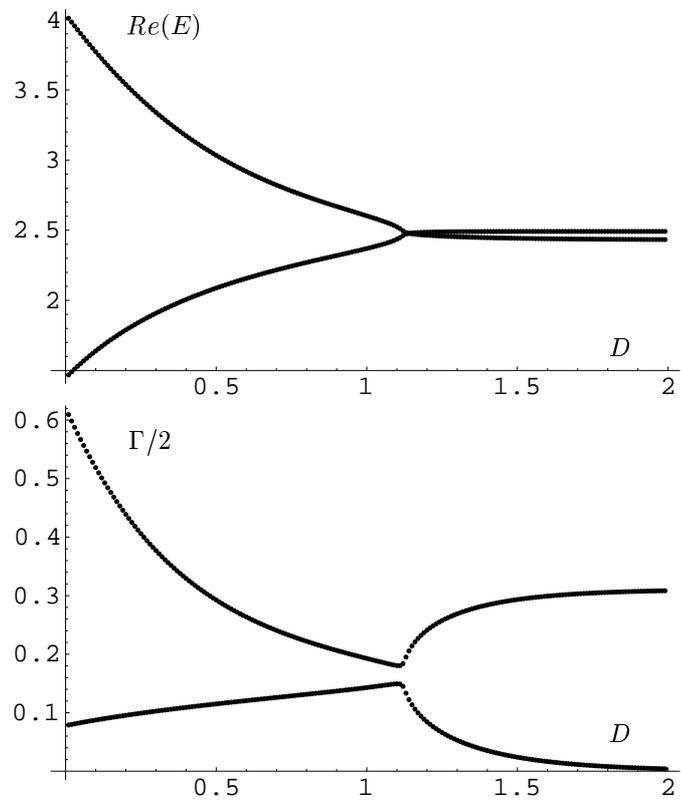
\end{document}